# VoIP Technology: Security Issues Analysis

**Amor Lazzez**

Taif University, Kingdom of Saudi Arabia
*a.lazzez@gmail.com*; *a.lazzez@tu.edu.sa*

**Abstract:** *Voice over IP (VoIP) is the technology allowing voice and multimedia transmissions as data packets over a private or a public IP network. Thanks to the benefits that it may provide, the VoIP technology is increasingly attracting attention and interest in the industry. Actually, VoIP allows significant benefits for customers and communication services providers such as cost savings, rich media service, phone and service portability, mobility, and the integration with other applications. Nevertheless, the deployment of the VoIP technology encounters many challenges such as architecture complexity, interoperability issues, QoS issues, and security concerns. Among these disadvantages, VoIP security issues are becoming more serious because traditional security devices, protocols, and architectures cannot adequately protect VoIP systems from recent intelligent attacks. The aim of this paper is carry out a deep analysis of the security concerns of the VoIP technology. Firstly, we present a brief overview about the VoIP technology. Then, we discuss security attacks and vulnerabilities related to VoIP protocols and devices. After that, we talk about the security profiles of the VoIP protocols, and we present the main security components designed to help the deployment of a reliable and secured VoIP systems.*

**Keywords:** VoIP, Vulnerabilities, Security Attacks, Security Mechanisms

## 1. INTRODUCTION

Voice over IP (VoIP) [1, 2] has been prevailing in the telecommunication world since its emergence in the late 90s, as a new technology transporting multimedia over the IP network. The reason for its prevalence is that, compared to legacy phone system, VoIP allows significant benefits such as cost savings, the provision of new media services, phone portability, and the integration with other applications [1, 2, 3]. Despite these advantages, the VoIP technology suffers from many hurdles such as architecture complexity, interoperability issues, QoS concerns, and security issues [1]. Among these disadvantages, VoIP security issues are becoming more serious because traditional security devices, protocols, and architectures cannot adequately protect VoIP systems from recent security attacks.

VoIP technology is characterized by a set of vulnerabilities, meaning the flaws that may be exploited by an attacker to perform security attacks. There are two types of vulnerability in VoIP [1, 2]. One is the inherited vulnerability coming from the infrastructure (network, operating system, etc.) that VoIP applications are running on. The other is its own vulnerability coming from VoIP protocols and devices. All components involved in the deployment of VoIP service have vulnerable elements that affect it directly or indirectly. The main vulnerable components in a VoIP system are the operating system of the VoIP application, the VoIP application itself, the VoIP protocols, the management interface, and the network devices (switch, router).

VoIP vulnerabilities can be exploited to create many different kinds of attacks. Security attacks can be categorized as four different types: attacks against availability, confidentiality, integrity, and social context [1, 2]. An attack against availability aims at VoIP service interruption, typically, in the form of Denial of Service (DoS). Call flooding, malformed messages, spoofed messages, and call hijacking are examples of attacks against service availability. Unlike attacks against availability, attacks against confidentiality do not impact current communications generally, but provide an unauthorized means of capturing media, identities, patterns, and credentials that are used for subsequent unauthorized connections or other deceptive practices. Eavesdropping media, call pattern tracking, data mining, and reconstruction are attacks against confidentiality. An attack against integrity alters transmitted data or signaling traffic after being intercepted in the middle of the network. The alteration may consist of deleting, injecting, or replacing certain information in the transmitted VoIP traffic. The typical examples are call rerouting, call black holing, media injection, and media degrading. An attack against social context focuses on how to manipulate the social context between communication parties so that an attacker can misrepresent himself as a trusted entity and convey false information to the target user. The typical examples are misrepresentation (identity, authority, rights, and content), voice spam, instant message spam, presence spam, and phishing.

To prevent these attacks, and hence help the deployment of secured VoIP systems, VoIP signaling and media transmission protocols (SIP, H.323, IAX, and RTP) define specific security mechanisms as part of the protocols, or recommend combined solution with other security protocols (IPSec, SRTP, etc.) [1,2]. Actually, H.235 [1, 3] provides authentication, privacy and integrity for H-323 framework [1, 2, 3]. SIP protocol [1, 5] describes several security features such as massage authentication, message encryption, media encryption, and network layer security. The IAX VoIP framework [6, 7, 8] allows message authentication and confidentiality, and supports NAT (Network Address Translation) and firewall traversal. In addition to the security capabilities of the VoIP protocols, specific security devices have been designed to enhance the security of VoIP systems [1, 2]. The security devices are primarily designed for providing security services like access control, intrusion detection, DoS protection, lawful interception, and so on. Examples of those devices are

VoIP-aware firewall, NAT, and SBC (Session Border Controller). The use of a single security device or mechanism cannot be sufficient to perfectly protect a VoIP system. Hence, a security policy incorporating different security schemes should be designed based on the analysis of the VoIP system vulnerabilities to ensure a perfect system security.

The remaining of this paper is organized as follows. Section 2 presents an overview about VoIP architectures and protocols. First, we present the VoIP architectures. Then, we highlight the benefits leading to the ever-growing of the VoIP popularity. After that, we present a brief overview about the main VoIP protocols. Section 3 focuses on the actual VoIP vulnerabilities, meaning the flaws that may be exploited by an attacker to perform a security attacks. Section 4 presents an overview about the VoIP security attacks. We present a classification of the VoIP attacks into four categories based on the infected security service (availability, integrity, confidentiality, and authentication), as well as typical attack examples of each category. To prevent VoIP security attacks, VoIP protocols define specific security mechanisms as part of the protocols, or recommend combined solution with other security protocols which help the deployment of secured VoIP systems. Section 5 presents the security capabilities of the main VoIP protocols. In addition to the security capabilities of the VoIP protocols, specific security devices have been designed to enhance the security of VoIP systems. Section 6 presents the main VoIP security devices and shows the security potential of each device. Section 7 concludes the paper.

## 2. BASICS OF VoIP TECHNOLOGY

VoIP is a rapidly growing technology that delivers voice communications over Internet or a private IP network instead of the traditional telephone lines [1-4]. VoIP involves sending voice information in the form of discrete IP packets sent over Internet rather than an analog signal sent throughout the traditional telephone network. VoIP helps the provision of significant benefits for users, companies, and service providers. Cost savings, the provision of new communication services, phone and service portability, mobility, and the integration with other applications are examples of the VoIP benefits. Yet, the deployment of the VoIP technology encounters many difficulties such as architecture complexity, interoperability issues, QoS issues, and security concerns [1-4]. One of the main features of the VoIP technology is that it may be deployed using a centralized or a distributed architecture. Even though they are currently widely used, client-server VoIP systems suffer from many hurdles [2, 4]. In order to overcome the shortcomings of the client-server model, the development community starts tending towards the deployment of the VoIP service using a peer-to-peer decentralized architecture [2, 9, 12].

In the following subsections, we first we highlight the benefits of the VoIP technology leading to the ever-increasing of its popularity. Then, we present the main architectures used in the deployment of the VoIP technology. After that, we present a brief overview the most important VoIP protocols. Finally, we mention the main concerns of the VoIP technology.

### 3.1 VoIP Benefits

The key benefits of the VoIP technology are as follows [1, 3, 4]:

*Cost savings:* The most attractive feature of VoIP is its cost-saving potential. Actually, for users, VoIP makes long-distance phone calls inexpensive. For companies, VoIP reduces cost for equipment, lines, manpower, and maintenance. For service providers, VoIP allows the use of the same communication infrastructure for the provision of different services which reduces the cost of services deployment.

*Provision of new communication services:* In addition to the basic communications services (phone, fax), the VoIP technology allows users to check out friends' presence (such as online, offline, busy), send instant messages, make voice or video calls, and transfer images, and so on.

*Phone portability:* VoIP provides number mobility; the phone device can use the same number virtually everywhere as long as it has proper IP connectivity. Many businesspeople today bring their IP phones or soft-phones when traveling, and use the same numbers everywhere.

*Service mobility:* Wherever the user (phone) goes, the same services will be available, such as call features, voicemail access, call logs, security features, service policy, and so on.

*Integration and collaboration with other applications:* VoIP allows the integration and collaboration with other applications such as email, web browser, instant messenger, social-networking applications, and so on.

### 3.2 VoIP Architecture

One of the main features of the VoIP technology is that it may be deployed using a centralized or a distributed architecture [2, 4]. The majority of current VoIP systems are deployed using a client-server centralized architecture. A client-server VoIP system relies on the use of a set of interconnected central servers known as gatekeepers, proxy servers, or soft-switches. The central servers are responsible for users' registration as well as the establishment of VoIP sessions between registered users. Figure 1 shows an example of a VoIP system deployed using the client-server architecture. As it is illustrated in the figure, each central server handles (registers, establishes a session with a local or a distant user, etc.) a

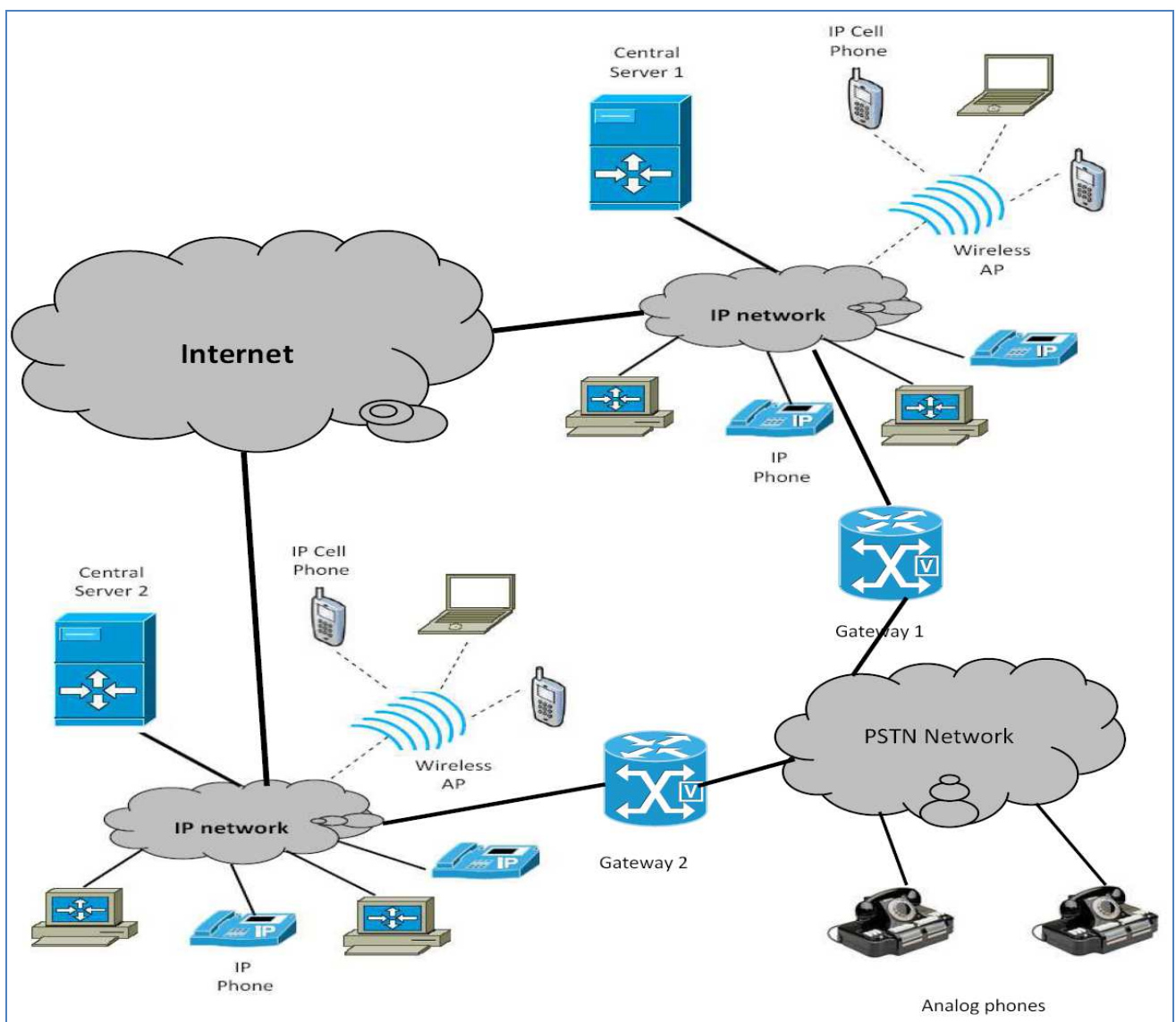
Figure1: Client-Server VoIP Architecture: An illustrative example

set of users. Each user must be registered on one of the central servers (registrar server) to be able to exchange data with other registered users. A user gets access to the service only over the registrar server.

Even though they are currently widely used, client-server VoIP systems suffer from the following many hurdles. The main issues of the client-server VoIP systems are the presence of single points of failure (central servers), scalability, availability, and security [11]. In order to overcome the shortcomings of the client-server model, and help the development of scalable and reliable VoIP systems, the development communities start tending towards the deployment of the VoIP service using a peer-to-peer decentralized architecture. Actually, a peer-to-peer VoIP system [2, 9, 12] allows service provision through the establishment of a symmetric collaboration between the system nodes (peers) communicating according to a given logic architecture (overlay). This helps the deployment of scalable, cost-effective, and more reliable systems VoIP systems.

### 3.3 VoIP Protocols

The deployment of any multimedia application such as VoIP, videoconference, or network gaming requires a signaling protocol to set up sessions between end points, and a distinct protocol to transmit the media streams. The standard protocol used to exchange media streams between the endpoints of an established session.

#### 3.3.1 VoIP Media Transport Protocols

The majority of VoIP systems rely on the use of the Real-Time Transport Protocol (RTP) for data transmission during a VoIP session. Secure RTP (SRTP) has been recently proposed by the IETF as a secured version of the RTP protocol.

*RTP Protocol:* Defined in RFC 3550, RTP protocol defines a standardized packet format for delivering audio and video over IP networks [1-4]. RTP is used in conjunction with the RTP Control Protocol (RTCP). While RTP carries the media streams (audio and video), RTCP monitors transmission statistics and the provided QoS and aids synchronization of multiple streams.

*SRTP Protocol:* SRTP protocol defines a security profile of RTP, intended to provide the authentication, the confidentiality, and the integrity of RTP messages [1, 2]. Since RTP is used in conjunction with RTCP, SRTP is closely related to SRTCP (Secure RTCP) which is used to control the SRTP session.

### 3.3.2 VoIP Signaling protocols

Given that the majority of current VoIP systems are deployed using a client-server centralized architecture, in the subsequent, we only consider the main signaling protocols used for the deployment of client-server VoIP systems; H323, SIP, and IAX.

*H323:* Standardized by the International Telecommunication Union (ITU), H323 [4] is the first signaling approach publicly used for the deployment of VoIP systems in conjunction with RTP protocol. H323 standard encompasses many protocols such as H225, H245, and H235. H.225 defines call setup messages and procedures used to establish a call, as well as messages and procedures used for users registration, and call admission. H.245 defines control messages and procedures used to exchange communication capabilities such as the supported codec. H235defines security profiles for H.323, such as authentication, message integrity, signature security, and voice encryption.

*SIP:* Allowing system flexibility and security, SIP is nowadays the most used VoIP signaling protocol [5]. SIP is an application layer protocol that works in conjunction with several other application layer protocols that identify and carry the session media. Media identification and negotiation is achieved with the Session Description Protocol (SDP). Media streams (voice, video) are transmitted using RTP protocol which may be secured by the SRTP protocol. For secure transmissions of SIP messages the Transport Layer Security (TLS) may be used. SIP also provides a suite of security services including DoS prevention, authentication, integrity, and confidentiality.

*IAX:* Currently, IAX (Inter-Asterisk Exchange) is one of the most used approaches for the deployment of VoIP systems [6-8]. In contrast with H323 and SIP protocols which are limited to signaling tasks, IAX protocol ensures both signaling and media transmission in an IAX-based VoIP system. IAX provides a suite of security services. Actually, it allows message authentication and confidentiality, and supports NAT traversal.

### 3.4 VoIP Disadvantages

Even though it allows significant benefits, the VoIP technology suffers from many hurdles [1, 2]. In the following a brief presentation of the main VoIP disadvantages.

*Complicated service and network architecture:* the integration of different services (voice, video, data, and so on) into the same network makes it difficult the design of the network architecture because different protocols and devices are involved for each service, and various characteristics are considered for each media. It also causes various errors and makes it harder to troubleshoot and isolate them.

*Interoperability issues between different applications, or products:* Different protocols (H323, SIP, IAX, and MGCP) have been proposed for the deployment of VoIP systems. This leads to an interoperability issues between the VoIP devices developed based on different protocols. Interoperability issues still come up between products using the same protocol due to the multitude of protocol versions, and the ways of implementation.

*Quality of service (QoS) issues:* The QoS aspect was not much considered when the IP technology was designed. That is why, the IP technology remains inefficient to support traffic with different QoS constraints despite the development of different approaches (DiffServ, IntServ) for the enhancement of the QoS provided by an IP network.

*Security issues:* In the legacy phone system (PSTN: Public Switched Telephone Network), the main security issue is the interception of conversations that require physical access to phone lines. In VoIP security issues are much more than that. Actually, in VoIP systems many elements (IP phones, access devices, media gateways, proxy servers, and protocols) are involved in setting up a call and transferring media between two endpoints. Each element has vulnerable factors that may be exploited by attackers to carry out security attacks.

Among the above presented disadvantages, VoIP security issues are becoming more serious because traditional security devices, protocols, and architectures cannot adequately protect VoIP systems from recent intelligent attacks. In the following sections, we present a deep analysis of the security issues of the VoIP technology. First, we present the main vulnerabilities of the VoIP systems. Then, we show how these vulnerabilities may be exploited by hackers to carry out different kinds of security attacks. Finally, we discuss how we can secure a VoIP system against security attacks.

## 4 VULNERABILITIES OF VOIP SYSTEMS

In system and network security, vulnerability is a flaw or a weakness that may be exploited by an attacker to carry out a security attack. VoIP has two types of vulnerability [1, 2]. The first one is the inherited vulnerability which comes from the infrastructure (network, operating system, web server, and so on) used for the deployment of VoIP applications. The other is the vulnerability coming from VoIP protocols and devices, such as IP phone, voice gateway, media server, signaling controller, etc.

The following subsections present the sources of vulnerabilities as well as the vulnerable components in a VoIP system.

### 4.1 Sources of Vulnerabilities

*IP-Based Network Infrastructure:* As the name VoIP implies, all traffic flows over IP networks and inherits the vulnerability of IP networks, such as malicious IP fragmentation, network viruses, or worms.

*Public Networks:* In most cases, VoIP traffic is transmitted over Internet where anonymous people including hackers may send and receive traffic.

*Open VoIP Protocol:* Most VoIP protocols, such as SIP or H.323, are standardized and open to the public. Hence, an attacker can create malicious client or server program based on the protocol specification in order to get access to a target VoIP servers or clients. Moreover, the openness of a VoIP protocol helps malicious people to identify and take advantage of its vulnerabilities.

*Voice and Data Integration:* Even though it allows significant benefits, the integration of voice and regular data traffic in the same network results into new traffic engineering issues. Actually, the integration of traffic with different QoS and security requirements makes the traffic engineering tasks (securing, switching, queuing, and s on) more complex and difficult.

*Lack of Specific Security Mechanisms:* While many data security mechanisms like firewalls may enhance the security of VoIP systems, it is still not enough to protect VoIP systems from today's malignant attacks.

*Real-Time Media Transfer:* Unlike common communication services like email, VoIP service requires a real-time transfer of media traffic which involves hard QoS constraints in terms of packet delay, and packet delay variation (jitter). Hence, minor packet delay or jitter could be recognized by users and impact the overall QoS. An attacker may overload the VoIP network (Calla flooding for example) to affect the provided QoS, and thus the system reliability.

*Exposed Interface:* The majority of current VoIP systems are deployed using a client-server architecture. Even though, VoIP servers are located in a protected network, the interface modules receiving call requests are open to clients that are located in an open or public network. This allows attackers to perform a ports scan to find out the exposed interface modules, and then carry out a security attack (DoS for example) by sending malicious traffic.

*Endpoints Mobility:* The PSTN (Public Switched Telephone Network) phone system assigns a dedicated phone line to a certain number. Thus, an attacker requires physical access to spoof the identity (the telephone number or line) of a regular user of the PSTN phone system. Unlike PSTN technology, VoIP phone systems allow endpoints mobility, which makes the protection against identity spoofing harder.

### 4.2 Vulnerable Components

In the following of this subsection, we present subsequent, a brief overview of the main vulnerable components involved in the deployment of a regular VoIP system [1, 2].

*Operating system:* VoIP applications are affected by the vulnerabilities of the operating systems are running on. The frequent security patches for the regular operating systems (Windows, Unix, Lunix) prove that they always have vulnerabilities.

*VoIP application:* A VoIP application (Skype, Google Talk, etc.) itself may have security issues because of bugs or errors, which could make VoIP service insecure.

*VoIP protocols:* The deployment of a VoIP application involves a signaling protocol (H323, SIP, IAX), and a media transmission protocol (RTP, RTCP). These protocols are vulnerable to different kinds of attacks which may affect the VoIP service provided based on these protocols.

*Management interface:* For management purposes, the majority of VoIP devices have different service interfaces such as SNMP, SSH, Telnet, and HTTP. A service interface may be a source of vulnerability, especially when being configured carelessly. For example, if a VoIP device uses the default ID/password for its management interface, it is easy for an attacker to break in.

*TFTP Server:* Many VoIP devices download their configurations from a TFTP server. An attacker could impersonate a TFTP server by spoofing the connection, and then distribute a malicious configuration to the VoIP equipment.

*Access device (switch, router):* All VoIP traffic flows through access devices (switch, router) that are in charge of switching or routing. Compromised access devices could create serious security issues because they have full control of packets.

*Network:* VoIP traffic is affected by the vulnerabilities of the IP network through which it is transmitted. An IP network vulnerability may be due to a bad configuration of a network device (switch, router, firewall, etc.) or a bug in one of the involved protocols (IP, UDP, and so on).

## 5 VoIP Security Attacks

The VoIP vulnerabilities presented in the previous section may be exploited by hackers to carry out different kinds of security attacks. Attackers may disrupt media service by flooding traffic, collect privacy information by intercepting call signaling or call content, hijack calls by impersonating servers or impersonating users, make fraudulent calls by spoofing identities, and so on.

There are many possible ways to categorize the security attacks. The first version of the IETF draft classified the security attacks into the following four categories: Interception and modification attacks, Interruption-of-service attacks, abuse-of-service attacks, and social attacks [13]. In [2], the authors consider the following categories of VoIP security attacks: service disruption and annoyance, eavesdropping and traffic analysis, masquerading and impersonation, unauthorized access, and fraud. In [1], the author classifies the security attacks into four categories as follows: attacks against availability, attacks against confidentiality, attacks against integrity, and attacks against social context.

In the following of this section, we present a brief overview about the main VoIP attacks according to the taxonomy presented in [1], which we adopt as it is the newest presented taxonomy compared to the other listed ones.

## 5.1 Attacks against availability

Attacks against availability aim at VoIP service interruption, typically in the form of Denial of Service (DoS). The main attacks against availability are: call flooding, malformed messages, spoofed messages, call hijacking, server impersonating, and Quality of Service (QoS) abuse. In the following, a brief presentation of these attacks.

*Call Flooding:* an attacker floods valid or invalid heavy traffic (signals or media) to a target system (for example, VoIP server, client, and underlying infrastructure) which breaks down the system or drops its performance significantly.

*Malformed Messages:* An attacker may create and send malformed messages to the target server or client for the purpose of service interruption. A malformed message is a protocol message with wrong syntax. The server receiving this kind of unexpected message could be confused (fuzzed) and react in many different ways depending on the implementation. The typical impacts are as follows: infinite loop, buffer overflow, inability to process other normal messages, and system crash.

*Spoofed Messages:* An attacker may insert fake (spoofed) messages into a certain VoIP session to interrupt the service, or steal the session. The typical example is call teardown. For this example, the attacker creates and sends a call termination message (for example SIP Bye) to a communicating device to tear down a call session. This attack requires the stealing of session information (Call-ID) as a preliminary.

*Call Hijacking:* Hijacking occurs when some transactions between a VoIP endpoint and the network are taken over by an attacker. The transactions can be a registration, a call setup, a media flow, and so on. This hijacking can make serious service interruption by disabling legitimate users to use the VoIP service. It is similar to call teardown in terms of stealing session information as a preliminary, but the actual form of attack and impact are different. The typical examples are registration hijacking, and media session hijacking.

*QoS Abuse:* The elements of a media session are negotiated between VoIP endpoints during call setup time, such as media type, coder-decoder (codec) bit rate, and payload type. An attacker may intervene in this negotiation and abuse the Quality of Service (QoS), by replacing, deleting, or modifying codecs or payload type. Another method of QoS abuse is exhausting the limited bandwidth with a malicious tool so that legitimate users cannot use bandwidth for their service.

## 5.2 Attacks Against Confidentiality

Attacks against confidentiality provide an unauthorized means of capturing media, identities, patterns, and credentials that are used for subsequent unauthorized connections or other deceptive practices. The main types of confidentiality attacks are eavesdropping media, call pattern tracking, data mining, and reconstruction.

*Media Eavesdropping:* An unauthorized access to media packets. Two typical methods are used by attackers. One consists to compromise an access device (layer 2 switch for example) and duplicate the target media to an attacker's device. The other way is that an attacker taps the same path as the media itself, which is similar to legacy tapping technique on PSTN. For example, the attacker may get access to the T1 itself and physically splits the T1 into two signals.

*Call Pattern Tracking:* Call pattern tracking is the unauthorized analysis of VoIP traffic from or to any specific nodes or network so that an attacker may find a potential target device, access information (IP/port), protocol, or vulnerability of network. It could also be useful for traffic analysis; knowing who called who, and when.

*Data Mining:* The general meaning of data mining in VoIP is the unauthorized collection of identifiers that could be user name, phone number, password, URL, email address, strings or any other identifiers that represent phones, server nodes, parties, or organizations on the network. These information may be used by an attacker for subsequent unauthorized connections such as service interruptions, confidentiality attacks, spam calls, etc.

## 5.3 Attacks Against Integrity

Attack against integrity consists in the alteration of the exchanged traffic (signaling messages or media packets) after intercepting them in the middle of the network. The alteration can consist of deleting, injecting, or replacing certain information in the VoIP message or media. Call rerouting and black holing are typical examples of attacks against the integrity of the signaling traffic. Media injection and degrading are examples of media integrity attacks.

*Call Rerouting:* An unauthorized change of call direction by altering the routing information in the signaling message. The result of call rerouting is either to exclude legitimate entities or to include illegitimate entities in the path of call signal or media.

*Media injection:* An unauthorized method in which an attacker injects new media into an active media channel. The consequence of media injection is that the end user (victim) may hear advertisement, noise, or silence in the middle of conversation.

*Media degrading:* An unauthorized method in which an attacker manipulates media or media control packets relative to an established communication session in order to reduce the quality of data communication (QoS). For instance, an attacker intercepts RTCP packets in the middle, and changes the sequence number of the packets so that the endpoint device may play the media with wrong sequence, which degrades the quality.

## 5.4 Attacks Against Social Context

An attack against social context focuses on how to manipulate the social context between communicating entities so that an attacker can misrepresent himself as a trusted entity and convey false information to the target user (victim). The typical attacks against social context are misrepresentation of identity, authority, rights, and content, spam of call and presence, and phishing.

*Misrepresentation:* It corresponds to the intentional presentation of a false identity, authority, rights, or content as if it were true so that the target user (victim) or system may be deceived by the false information. Identity misrepresentation is the method of presenting an identity with false information, such as false caller name, organization, email address, or presence information. Authority or rights misrepresentation is the method of presenting false information to an authentication system to obtain the access permit, or bypassing an authentication system. Content misrepresentation is the method of presenting false content as if it came from a trusted source of origin. It includes false impersonation of voice, video, text, or image of a caller.

*Spam:* Call spam is defined as a bulk unsolicited set of session initiation attempts (INVITE requests), attempting to establish a voice or video communications session. If the user should answer, the spammer proceeds to relay their message over real-time media. Presence spam is defined as a bulk unsolicited set of presence requests (for example, SIP SUBSCRIBE requests) in an attempt to get on the "buddy list" of a user to subsequently carry out a call spam (INVITE request).

*Phishing:* An illegal attempt to obtain somebody's personal information (for example, ID, password, bank account number, credit card information) by posing as a trust entity in the communication. The typical method is that an attacker picks target users and creates request messages (SIP INVITE for example) with spoofed identities, pretending to be a trusted party. When the target user accepts the call request, the phisher provides fake information (for example, bank policy announcement) and asks for personal information. Some information like user name and password may not be directly valuable to the phisher, but it may be used to access more information useful in identity theft.

## 6 SECURITY ABILITIES OF VOIP PROTOCOLS

To prevent the above presented attacks, and hence help the deployment of secured VoIP systems, VoIP protocols (SIP, H.323, IAX) define specific security mechanisms as part of the protocols, or recommend combined solution with other security protocols (IPSec, SRTP, etc.) [1, 2]. In the following subsections, we present a brief overview about the security abilities of the dominating protocols in the current VoIP systems: H323, SIP, and IAX for signaling and RTP/RTCP for media transport.

### 6.1 H.323 Security Abilities

Security for H.323 is described by the ITU-T standard H235"Security and Encryption for H-Series Multimedia Terminals" [1, 2, 4]. The scope of this standard is to provide authentication, privacy and integrity for H-323. Different profiles have been defined for the use of the H235 security protocol. Each profile is defined by a specific annex. Annex D describes a simple, password-based security profile. Annex E describes a profile using digital certificates and dependent on a fully-deployed public-key infrastructure. Annex F combines features of both annex D and annex E.

*Annex D:* Defines a simple, baseline security profile. The profile provides basic security by simple means, using secure password-based cryptographic techniques. This profile is applicable in an environment where a password/symmetric key may be assigned to each H.323 entity (terminal, gatekeeper, gateway, or MCU). It provides authentication and integrity for H.225 protocols (RAS, and Q931), and tunneled H.245 using password-based HMAC-SHA1-96 hash. Optionally, the voice-encryption security profile can be combined smoothly with the baseline security profile. Audio streams may be encrypted using the voice-encryption security profile deploying Data Encryption Standard (DES), RC2-compatible or triple-DES, and using the authenticated Diffie-Hellman key-exchange procedure.

*Annex E:* Describes a security profile deploying digital signatures that is suggested as an option. H323 entities (terminals, gatekeepers, gateways, MCUs, and so on) may implement this signature security profile for improved security or whenever required. Typically, it is applicable in environments with potentially many terminals where password/symmetric key assignment is not feasible. The signature security profile overcomes the limitations of the simple, baseline security profile of Annex D.

*Annex F:* Describes an efficient and scalable, public key infrastructure (PKI)-based hybrid security profile deploying digital signatures from Annex E and deploying the baseline security profile from Annex D. With this security profile, digital signatures from the signature security profile in annex E are deployed only where absolutely necessary, and highly efficient symmetric security techniques from the baseline security profile in Annex D are used otherwise. The hybrid security profile overcomes the limitations of the simple, baseline security profile of Annex D as well as certain drawbacks of Annex E, such as the need for higher bandwidth and increased performance needs for processing, when strictly applied.

### 6.2 SIP Security Abilities

The SIP protocol describes several security features [1, 2]. The main security features of the SIP protocol are: message authentication, message encryption, media encryption, transport layer security, and network layer security. Only message authentication is ensured by SIP protocol, and the others abilities are allowed by other security protocols such as S/MIME, SRTP/SRTCP, TLS, and IPSec. In the following, a brief presentation of the main security features of the SIP signaling protocol.

*Message Authentication:* SIP ensures the authentication of signaling messages (REGISTER, INVITE, and BYE) to avoid registration hijacking attacks and prevent unauthorized calls and DoS or annoyance attacks.

*Message Encryption:* SIP relies on the S/MIME (Secure/Multipurpose Internet Mail Extensions) protocol to encrypt the headers of the signaling messages (except the "Via", and "Route" headers) which helps end-to-end confidentiality, integrity, and authentication between participants. S/MIME provides the flexibility for more granular protection of header information in SIP messages as it allows a selectively protection of SIP message fields.

*Media encryption:* SRTP (Secure RTP) protocol ensures the encryption of media packets encryption which helps the guarantee of the confidentiality and integrity of exchanged media. Section 5.4 details the security capabilities of SRTP protocol.

*Transport Layer Security (TLS):* TLS protocol is used to provide a transport-layer security of SIP messages (requests, responses). Actually TLS ensures the encryption of entire SIP requests and responses which ensures the confidentiality and integrity of messages.

*Network Layer Security:* SIP relies on the use of IPSec at the network layer which enhances the security of IP network communications by encrypting and authenticating data. IPSec is very useful to provide security between SIP entities, especially between a user agent (UA) and a proxy server.

### 6.3 IAX Security Abilities

As it is mentioned above, IAX allows message authentication and confidentiality, and supports NAT (Network Address Translation) and firewall traversal [6, 7, 8]. Actually, IAX protocol was deliberately designed to work behind firewalls and devices performing NAT. Moreover, IAX includes the ability to encrypt the streams between endpoints with the use of an exchanged RSA key, or dynamic key exchange at call setup, allowing the use of automatic key rollover.

### 6.4 RTP/RTCP Security Abilities

Secure RTP (or SRTP) [1, 2] defines a profile of RTP Protocol, intended to provide confidentiality, integrity, and authentication to media streams in both unicast and multicast applications. In addition to protecting the RTP packets, SRTP provides protection for the RTCP streams. The designers of SRTP focused on developing a protocol that can provide adequate protection for media streams but also maintain key properties to support wired and wireless networks in which bandwidth or underlying transport limitations may exist.

## 7 VOIP SECURITY DEVICES

In addition to the security capabilities of the VoIP protocols, specific security devices have been designed to enhance the security of VoIP systems [1, 2]. The security devices are primarily designed for providing security services like access control, intrusion detection, DoS protection, and so on. Examples of those devices are VoIP-aware firewall, Network Address Translation (NAT), and Session Border Controller (SBC).

### 7.1 VoIP-aware firewall

A firewall is a key security device in an IP network allowing the protection of the internal network from external attacks. The general function is to block certain types of traffic based on the source/destination IP address, the used transport protocol (TCP, UDP), the source/destination port number, the traffic direction (input, output), and the traffic type (RTP, HTTP, SMTPP). VoIP traffic may be handled using a regular or a VoIP-aware firewall. Compared to a regular firewall which handles packets only at the network and transport layer, a VoIP-aware firewall has the additional capability to inspect and manipulate VoIP packets at the application layer [1]. Actually, a VoIP-aware firewall allows:

*Inspection of protocol messages:* consists to check out the integrity of protocol messages (SIP messages), and blocks the originator if it detects any malformed messages.

*Protection against DoS attacks:* consists to detect any flooded messages and blocks the originator for a certain amount of time, based on a given policy. The policy may include number of call attempts per second, number of messages per second, number of invalid messages, etc.

*Control of the bandwidth utilization:* It can assign maximum bandwidth for each endpoint (or group), and block any overused endpoint.

### 7.2 Network Address Translation

Network Address Translation (NAT) [1, 2] is a method of connecting multiple computers to the Internet (or any other IP network) using one IP address. This allows home users and small businesses to connect their network to the Internet cheaply and efficiently. NAT automatically provides firewall-style protection without any special set-up. That is because it only allows connections that are originated on the inside network. This means, for example, that an internal client can connect to an outside FTP server, but an outside client will not be able to connect to an internal FTP server because it would have to originate the connection, and NAT will not allow that. It is still possible to make some internal servers available to the outside world via inbound mapping, which maps certain well know TCP ports (21 for FTP) to specific internal addresses, thus making services such as FTP or Web available in a controlled way.

### 7.3 Session Border Controller

Session Border Controller (SBC) [1, 2] is a controlling device located in a border of two network sessions. A network session may be an access network, a core network, and so on. For instance, from a VoIP service provider's perspective, there are two network borders. One is between the customer's access network and the core network (service provider's network). The other is between the core network and the other service provider's network (peer network). The role of a session border controller is to resolve border concerns that include interoperability and security issues. Security issues are mainly due to the exposure of a network session (a core network for example) to other network sessions (peer network, or a customer access network) which may help malicious users form a network session to attack resources (VoIP server, proxy, etc.) in another network session. However, interoperability issues are basically due to the interaction between network sessions using different devices and protocols.

## 8 CONCLUSION

In this paper, we have presented a deep analysis of the security concerns of the VoIP technology. Firstly, we have presented a brief overview about the basics of the VoIP technology. Then, we have discussed the security vulnerabilities and attacks related to VoIP protocols and devices. After that, we have presented the countermeasures that should be considered to help the deployment of secured VoIP systems. A future work will address another important issue in the deployment of VoIP technology; the ability to support the QoS constraints of the voice and video applications.

## AUTHOR

**Amor Lazzez** is currently an Assistant Professor of Computer and Information Science at Taif University, Kingdom of Saudi Arabia. He received the Engineering diploma with honors from the high school of computer sciences (ENSI), Tunisia, in June 1998, the Master degree in Telecommunication from the high school of communication (Sup'Com), Tunisia, in November 2002, and the Ph.D. degree in information and communications technologies form the high school of communication, Tunisia, in November 2007. Dr. Lazzez is a researcher at the Digital Security research unit, Sup'Com. His primary areas of research include design and analysis of architectures and protocols for optical networks.